\documentclass[prb, fleqn,twocolumn, showpacs, showkeys]{revtex4}

\usepackage{graphicx}
\usepackage{amsmath}

\input{tcilatex}

\begin{document}

\title{Anisotropy of the paramagnetic susceptibility in
LaTiO$_{3}$: The electron-distribution picture in the ground
state}

\author{R.~M.~Eremina$^{a,c}$, M.~V.~Eremin$^{b,c}$, S. V. Iglamov$^{b}$,
J.~Hemberger$^{c}$, H.-A.~Krug~von~Nidda$^{c \ast}$,
F.~Lichtenberg$^{d}$, A.~Loidl$^{c}$}

\affiliation{$^{a}$E. K. Zavoisky Physical Technical Institute,
420029 Kazan, Russia}

\affiliation{$^{b}$Kazan State University, 420008 Kazan, Russia}

\affiliation{$^{c}$Experimental Physics V, Electronic Correlations
and Magnetism, Institute of Physics, University of Augsburg, 86135
Augsburg, Germany}

\email{Hans-Albrecht.Krug@physik.uni-augsburg.de}

\affiliation{$^{d}$Experimental Physics VI, Electronic
Correlations and Magnetism, Institute of Physics, University of
Augsburg, 86135 Augsburg, Germany}

\date{\today}

\begin{abstract}
The energy-level scheme and wave functions of the titanium ions in
LaTiO$_{3}$ are calculated using crystal-field theory and
spin-orbit coupling. The theoretically derived temperature
dependence and anisotropy of the magnetic susceptibility agree
well with experimental data obtained in an untwinned single
crystal. The refined fitting procedure reveals an almost
isotropic molecular field and a temperature dependence of the van
Vleck susceptibility. The charge distribution of the
$3d$--electron on the Ti positions and the principle values of the
quadrupole moments are derived and agree with NMR data and recent
measurements of orbital momentum $\langle l \rangle$ and
crystal-field splitting. The low value of the ordered moment in
the antiferromagnetic phase is discussed.
\end{abstract}




\maketitle






\section{\label{sec:intro}Introduction}

In the physics of highly correlated electron systems the
electronic orbitals and their interactions are in the focus of
recent experimental and theoretical research, because the
orbitals play a key role in the coupling of charge and spin of the
electrons with the lattice. Transition-metal oxides, where the
shape and anisotropy of the $d$--electron orbitals determine the
fundamental electronic properties, provide a rich field for this
kind of investigations. For example the perovskite titanates
$A$TiO$_3$ (with $A$ = Y, La, or some trivalent rare-earth ion)
are known as realization of a Mott insulator. The $3d^1$
electronic configuration of Ti$^{3+}$ corresponds to an
effectively half-filled conduction band, where the on-site
Coulomb repulsion inhibits double occupation of the Ti--sites
resulting in an insulating ground state.\cite{Okimoto1995}
Although from their electronic configuration these titanates seem
to be quite simple model systems, their orbital properties still
have to be resolved especially in the case of LaTiO$_3$.

The debate on the orbital ground state of LaTiO$_3$ was triggered
by its unusual magnetic properties. Below the N\'eel temperature
$T_{\rm N}=146$~K,\cite{Lichtenberg1991} LaTiO$_3$ reveals a
slightly canted $G$--type antiferromagnetic structure with an
ordered moment of $0.46 \mu_{\rm B}$,\cite{Goral1983,Meijer1999}
which is strongly reduced as compared to the spin-only value of $1
\mu_{\rm B}$ and, hence, indicates a strong importance of the
spin-orbit coupling. On the other hand the nearly isotropic
spin-wave dispersion with a small gap of about 3~meV contradicts
a dominant spin-orbit coupling.\cite{Keimer2000}

This puzzling situation originates from the fact that the
orthorhombic GdFeO$_3$ structure of LaTiO$_3$ deviates only weakly
from the ideal cubic perovskite structure: The quasicubic crystal
field of the nearly ideal oxygen octahedron surrounding the
Ti$^{3+}$ ion splits the five orbital $3d$ levels into a lower
$t_{2g}$ triplet and an excited $e_g$ doublet. The single
electron occupies the lower $t_{2g}$ triplet and is Jahn-Teller
active.\cite{Jahn1937} In principle the Jahn-Teller effect is
expected to lift the remaining threefold degeneracy resulting in
a distortion of the oxygen octahedron in favor of one of the
three orbitals. However, the competing influence of spin-orbit
coupling cannot be neglected in the case of a single electron in
a $t_{2g}$ level, as has been outlined already by
Goodenough\cite{Goodenough1968} and by Kugel and
Khomskii.\cite{Kugel1982} It is important to note that, as long
as the orbital triplet remains degenerate, the exchange
interactions are inherently frustrated even in a cubic
lattice.\cite{Khomskii2003}

To promote possible physics of this degeneracy in LaTiO$_3$, an
orbital-liquid ground state has been
suggested.\cite{Khaliullin2000} Further detailed theoretical
studies\cite{Khaliullin2001,Kikoin2003} favoring the
orbital-liquid picture worked out that the frustration can be
resolved via an order-by-disorder mechanism giving rise to
magnetic spin order with disordered orbital states. The observed
spin-wave excitations were found to be in accord with this model.
In a different theoretical approach \cite{Mochizuki2001,
Mochizuki2003, Pavarini2004, Craco2003, Solovyev2004} the crystal
field of the La ions caused by the GdFeO$_3$--type distortion has
been shown to lift the degeneracy of the Ti--$t_{2g}$--orbitals
and to stabilize the antiferromagnetic $G$--type order. In
Ref.~\onlinecite{Mochizuki2001, Mochizuki2003, Pavarini2004} the
orbital-ground state was derived as approximately $3z^2_{111}-r^2
= (d_{xy}+d_{yz}+d_{zx})/\sqrt{3}$. However, Solovyev
\cite{Solovyev2004} has found that the Hartree-Fock approximation
alone fails to provide the description of the magnetic properties
of LaTiO$_3$ and YTiO$_3$.

Several recent experimental investigations strongly support the
existence of orbital order in LaTiO$_3$. Specific-heat,
electrical resistivity, thermal-expansion, and infrared
experiments\cite{Hemberger2003} exhibit anomalies near the N\'eel
temperature, which indicate significant structural changes and
have been interpreted in terms of the influence of orbital order
via magneto-elastic interactions. Transmission-electron microscopy
revealed small atomic displacements ascribed to a weak Jahn-Teller
distortion.\cite{Arao2002} Detailed x-ray and neutron-diffraction
studies\cite{Cwik2003} of crystal and magnetic structure revealed
an intrinsic distortion of the oxygen octahedra, which leads to a
large enough splitting of the Ti--$t_{2g}$ triplet state. The
remeasured magnetic moment $\mu = 0.57(5)\mu_{\rm B}$ turned out
to be slightly larger than determined before.\cite{Cwik2003} The
reexamination of the Ti nuclear magnetic resonance
spectra\cite{Kiyama2003} proves a large nuclear quadrupole
splitting, which is ascribed to a rather large quadrupole moment
of the $3d$ electrons at the Ti sites. This discarded the earlier
interpretation\cite{Itoh1999} of the NMR results in terms of
orbital degeneracy and clearly favored the orbital order.

In this communication we perform a detailed analysis of the
temperature dependence and anisotropy of the magnetic
susceptibility of LaTiO$_3$, which we obtained on an untwinned
single crystal. In an earlier publication \cite{Fritsch2002} it
was mentioned that the anisotropy observed in the paramagnetic
regime requires to include the spin-orbit coupling into the
crystal-field calculation. In the present analysis we develop
this approach and go beyond the Hartree-Fock
approximation.\cite{Solovyev2004} Besides the spin-orbit coupling
we are taking into account the Ti--O exchange as well. We will
show that the obtained orbital-order pattern is basically in
agreement with NMR data \cite{Kiyama2003} and allows to describe
consistently the temperature dependence and anisotropy of the
observed experimental susceptibility.

\section{\label{cef}Crystal field analysis}

In LaTiO$_{3}$ the Ti$^{3+}$ ions (electronic configuration
$3d^1$, spin $s=1/2$) are situated in slightly distorted octahedra
formed by the oxygen ions. The dominant cubic component of the
crystal field splits the five $3d$--electron states into a lower
triplet $t_{2g}$ and an upper doublet $e_{g}$. The low-symmetry
component of the crystal field is expected to be small with
respect to the cubic one and, therefore, one may be tempted to
analyze the magnetic susceptibility using the basis of the
$t_{2g}$ states with a fictitious orbital momentum $\tilde{l}=1$,
only.\cite{Abragam1970} However, this procedure is not convenient
for LaTiO$_{3}$ for the following reason: Indeed, the wave
functions of the fictitious momentum $\tilde{l}=1$ are defined in
a local coordinate system $(x,y,z)$ with its axes parallel to the
$C_{4}$ axes of the non distorted octahedra. In the real structure
of LaTiO$_{3}$, there are four different fragments TiO$_{6}$,
which are distorted and rotated with respect to each other, i.e.
the $\tilde{l}=1$ basis should be rotated correspondingly for
each of the four inequivalent octahedra. During these rotations
all $3d$--electron states are mixed. In this situation it is
preferable to stay in the crystallographic coordinate system
using the full basis of $3d$--electron states.

Thus, to determine the energy-level scheme of Ti$^{3+}$ in
LaTiO$_3$, we start from the Hamiltonian
\begin{equation}\label{Hamiltonian}
{\cal H}_{0}=\xi(\mathbf{ls})+ {\underset{k=2;4}{\sum}}\,
\overset{k}{\underset{q=-k}{\sum}}
B_{q}^{(k)}C_{q}^{(k)}(\vartheta,\varphi).
\end{equation}
The first term denotes the spin-orbit coupling with spin
$\mathbf{s}$ and orbital momentum $\mathbf{l}$. For Ti$^{3+}$ the
parameter of the spin-orbit coupling is expected to be about $\xi
\approx 200$~K.\cite{Abragam1970} The second term represents the
crystal field with the spherical tensor
$C_{q}^{(k)}(\vartheta,\varphi)=\sqrt{2\pi/(2k+1)}
Y_{q}^{(k)}(\vartheta,\varphi)$. The crystal-field parameters
\begin{equation}
B_{q}^{(k)}=\underset{j}{\sum}
a^{(k)}(R_{j})(-1)^{q}C_{-q}^{(k)}(\vartheta_{j},\varphi_{j})
\end{equation}
are calculated using available data about the crystal structure.
\cite{table1996,MacLean1979,Eitel1986} The sum runs over the
lattice sites $R_{j}$.

The main contributions to the quantities $B_{q}^{(k)}$ originate
from the point charges $Z_{j}$ of the lattice and so called
exchange charges. Hence, the intrinsic parameters of the crystal
field are given by
\begin{equation}
a^{(k)}(R_{j})=-\frac{Z_{j}e^{2}\langle r^{k}
\rangle}{R_{j}^{k+1}} + a_{\rm ex}^{(k)}(R_{j}).
\end{equation}
The exchange contribution originates from the charge transfer
from oxygen into the unfilled $3d$--shell, i.e. the covalence
effect, and the direct titanium-oxygen exchange
coupling:\cite{Malkin1987, Eremin1977}
\begin{eqnarray}
a_{\rm ex}^{(2)}(R_{j})=\frac{G}{R_{j}}(S_{3d\sigma}^{2} +
S_{3ds}^{2}+S_{3d\pi }^{2})\nonumber \\
a_{\rm ex}^{(4)}(R_{j})=\frac{9G}{5R_{j}}(S_{3d\sigma}^{2} +
S_{3ds}^{2}-\frac{4}{3}S_{3d\pi }^{2}).
\end{eqnarray}
Here $S_{3d\sigma}$, $S_{3d\pi}$, and $S_{3ds}$ denote the
overlap integrals for Ti$^{3+}(3d^{1})$--O$^{2-}(2s^{2}2p^{6})$,
which are determined in local coordinate systems with the
$z$--axis along the titanium-oxygen bond. All integrals are
calculated using the Hartree-Fock wave
functions\cite{Clementi1964} of Ti$^{3+}$ and O$^{2-}$. The
parameter $G=7.2$ is an adjustable parameter, which we have
extracted from the cubic crystal-field splitting parameter
$10Dq$, which can be assumed as approximately similar for all
titanium oxides, as e. g. for Ti$^{3+}$ in Al$_{2}$O$_{3}$ with
$10Dq=19000$~cm$^{-1}$.\cite{Abragam1970}

In LaTiO$_{3}$ there is no inversion symmetry at the oxygen
position and, therefore, each oxygen ion exhibits a dipole moment
$\mathbf{d}_{i}=\alpha \mathbf{E}_{i}$, where $\alpha$ denotes
the polarization constant \cite{Faucher1982} and $\mathbf{E}_{i}$
is the electric field of the surrounding ions at the oxygen site
with number $i$. For the oxygen positions
\cite{MacLean1979,Eitel1986} O1 ($X=0.49036$, 0.25, $Z=0.07813$)
and O2 ($x=0.29144$, $y=0.04116$, $z=0.71036$) at $T=298$~K, the
values of the dipole moments (in units of $e$\AA) were calculated
as $d_x=-0.093$, $d_y=0$, $d_z=-0.001$ (O1) and $d_x=0.036$,
$d_y=0.018$, $d_z=0.037$ (O2), respectively. The relative signs
for the other three O1 and seven O2 positions change like the
signs of the corresponding coordinates ($X$, $Z$, and $x$, $y$,
$z$), e. g. for the O1 position ($X+0.5$, 0.25, $0.5-Z$) we
obtain $d_x=-0.093$, $d_y=0$, $d_z=0.001$ and so on. The
corresponding expressions for corrections to the crystal-field
parameters $B_{0}^{(2)},B_{2}^{(2) },B_{1}^{(2)}$ are calculated
as usual.\cite{Faucher1982}

In the crystallographic coordinate system, with the Cartesian axes
$x$, $y$, and $z$ chosen along the crystal axes $a = 5.6071$~\AA,
$b = 7.9175$~\AA, and $c = 5.6247$~\AA~in Pnma representation
(corresponding to $b$, $c$, and $a$ in Pbnm representation, which
is used in many papers), respectively, (values at room
temperature 298~K) we obtain the crystal-field parameters (in K)
for the titanium ion in position Ti1$(\frac{1}{2},\frac{1}{2},0)$
as given in Table~\ref{CFP}. For the other three titanium
positions the absolute values of $B_{q}^{(k)}$ are the same, but
their signs are different (cf. Table~\ref{signs}). Note that the
quantum mechanical contributions are comparable to the classical
ones and even dominate for $k=4$.

\begin{table}
\caption{Contributions to the crystal-field parameters in
LaTiO$_{3}$ at the Ti1 position $(\frac{1}{2}, \frac{1}{2}, 0)$
in units of K} \label{CFP}
\begin{tabular}{llll}
$B_{q}^{(k)}$ \hspace{0.5cm} & point charges \hspace{0.2cm} & exchange charges \hspace{0.2cm} & dipolar \\
$B_{0}^{(2)}$ & $1527$ & $720$ & $-819$ \\
$B_{1}^{(2)}$ & $-162-i376$ & $-301+i62$ & $1548-i413$ \\
$B_{2}^{(2)}$ & $-1229+i2496$ & $-941+i103$ & $430+i1525$ \\
$B_{0}^{(4)}$ & $-4486$ & $-7713$ & \hspace{0.3cm} small \\
$B_{1}^{(4)}$ & $-5828+i4105$ & $10951+i7733$ & \hspace{0.3cm} small \\
$B_{2}^{(4)}$ & $11325-i1699$ & $19452-i2160$ & \hspace{0.3cm} small \\
$B_{3}^{(4)}$ & $1827+i7634$  & $3407+i14371$ & \hspace{0.3cm} small \\
$B_{4}^{(4)}$ & $7638+i1713$  & $13047+i2963$ & \hspace{0.3cm} small \\
\end{tabular}
\end{table}

\begin{table}[tbp]
\caption{Relative signs of the parameters $B_{q}^{(k)}$ for Ti2,
Ti3, and Ti4 with respect to the signs for the Ti1 position in
LaTiO$_{3}$} \label{signs}
\begin{tabular}{llll}
\hspace{1.0cm} & Ti2$(0,\frac{1}{2},\frac{1}{2})$\hspace{0.5cm} &
Ti3$(\frac{1}{2},0,0)$\hspace{0.5cm} &
Ti4$(0,0,\frac{1}{2})$ \\
$B_{0}^{(2)}$ & + & + & + \\
$B_{1}^{(2)}$ & $\func{Re}-$, $\func{Im}-$ & $\func{Re}+$, $\func{Im}-$ & $\func{Re}-$, $\func{Im}+$ \\
$B_{2}^{(2)}$ & $\func{Re}+$, $\func{Im}+$ & $\func{Re}+$, $\func{Im}-$ & $\func{Re}+$, $\func{Im}-$ \\
$B_{0}^{(4)}$ & + & + & + \\
$B_{1}^{(4)}$ & $\func{Re}-$, $\func{Im}-$ & $\func{Re}+$, $\func{Im}-$ & $\func{Re}-$, $\func{Im}+$ \\
$B_{2}^{(4)}$ & $\func{Re}+$, $\func{Im}+$ & $\func{Re}+$, $\func{Im}-$ & $\func{Re}+$, $\func{Im}-$ \\
$B_{3}^{(4)}$ & $\func{Re}-$, $\func{Im}-$ & $\func{Re}+$, $\func{Im}-$ & $\func{Re}-$, $\func{Im}+$ \\
$B_{4}^{(4)}$ & $\func{Re}+$, $\func{Im}+$ & $\func{Re}+$, $\func{Im}-$ & $\func{Re}+$, $\func{Im}-$ \\
\end{tabular}
\end{table}

Using the crystal-field parameters listed above, for the position
Ti1$(\frac{1}{2},\frac{1}{2},0)$ we obtain the following energy
spectrum of five Kramers doublets with energies
$\varepsilon_{1,2}/k_{\rm B}=0$, $\varepsilon_{3,4}/k_{\rm
B}=2553$~K, $\varepsilon_{5,6}/k_{\rm B}=3214$~K,
$\varepsilon_{7,8}/k_{\rm B}=26773$~K, and
$\varepsilon_{9,10}/k_{\rm B}=27890$~K. This excitation spectrum
agrees perfectly with results from FIR experiments, which reveal
a hump in the optical conductivity close to
3000~K.\cite{Lunkenheimer2003} It is also in good agreement with
the results of recent spin-polarized photo-electron spectroscopy
experiments, which yield a crystal-field splitting of
0.12--0.30~eV, i.e. 1300--3300~K, of the $t_{2g}$
subshell.\cite{Haverkort2004} The corresponding wave functions in
$|m_{l},m_{s}\rangle$ quantization are written as follows:
\begin{eqnarray}
|\varepsilon_{n}\rangle &=&
\overset{+2}{\underset{m_l=-2}{\sum}}\hspace{0.3cm}\underset{m_s=\uparrow,\downarrow}{\sum}
a_{m_l,m_s}^{(n)}\mid m_{l},m_{s}\rangle.
\end{eqnarray}
In particular for one of the components of the ground doublet of
Ti1$(\frac{1}{2},\frac{1}{2},0)$ the coefficients are explicitly
given in Table~\ref{wave}. The other component of the ground state
can be obtained as Kramers conjugated state. Note that the
$g$-values $g_{z}=2\langle\varepsilon_{1}\mid
k_{z}l_{z}+2s_{z}\mid \varepsilon_{1}\rangle$,
$g_{x}=2\mid\langle\varepsilon_{1}\mid k_{x}l_{x}+2s_{x}\mid
\varepsilon_{2}\rangle\mid $, and $g_{y}=2\mid
\langle\varepsilon_{1}\mid k_{y}l_{y}+2s_{y}\mid
\varepsilon_{2}\rangle\mid$ are equal for all four titanium
positions, i.e. $g_{z}=1.81$, $g_{x}=1.73$, $g_{y}=1.79$, where
the reduction factors of the orbital momentum due to covalency
have been assumed as $k_{\alpha}=1$. The relatively small
deviation of the $g$ value from the spin-only value 2 displays
that the orbital momentum is rather small, again in agreement with
the recent spin-resolved photo-emission
experiments.\cite{Haverkort2004}

\begin{table}
\caption{Coefficients of the ground-state wave functions in
LaTiO$_{3}$ at the Ti1 position $(\frac{1}{2}, \frac{1}{2}, 0)$}
\label{wave}
\begin{tabular}{lll}
$a_{m_l,m_s}^{(1)}$ \hspace{0.5cm} & $m_s=\uparrow$ \hspace{1.0cm}& $m_s=\downarrow$\\
$m_{l} = 2$ & $-0.479-i0.191$ & $-0.033-i0.031$ \\
$m_{l} = 1$ & $0.136+i0.025$ & $0.005-i0.020$ \\
$m_{l} = 0$ & $-0.032+i0.608$ & $-0.011+i0.030$ \\
$m_{l} = -1$ & $0.154-i0.047$ & $-0.012-i0.007$ \\
$m_{l} = -2$ & $0.526-i0.186$ & $0.048$ \\
\end{tabular}
\end{table}

\begin{figure}
\includegraphics[width=6cm]{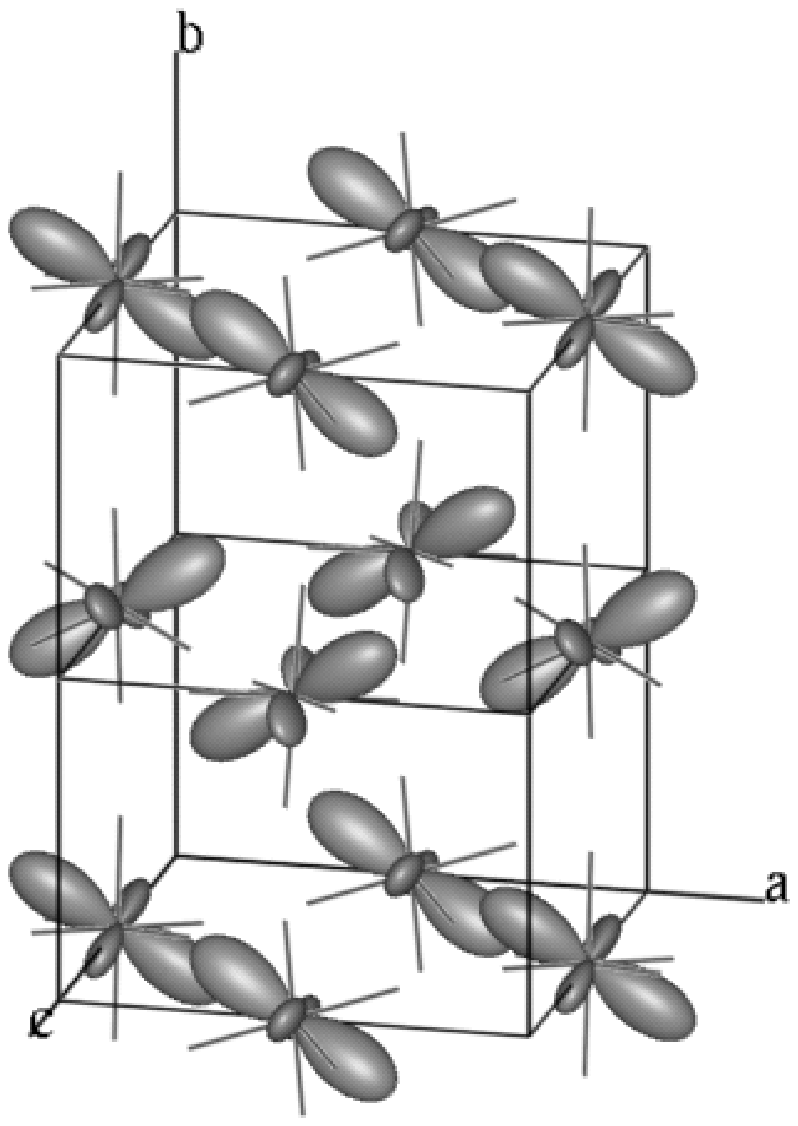}
\caption{Orbital order in LaTiO$_3$ as derived from the
crystal-field analysis\label{OO}}
\end{figure}

Figure~\ref{OO} illustrates the orbital order pattern due to the
derived ground-state wave function (cf. Table~\ref{wave}).
Basically, this is in agreement with the order patterns found by
Cwik {\it et al.},\cite{Cwik2003} by Kiyama and
Itoh,\cite{Kiyama2003} and by Pavarini {\it et
al.}\cite{Pavarini2004} However, in those works the wave functions
have been approximated in terms of the $t_{2g}^{(111)}$ basis
only, neglecting the spin-orbit coupling.

Having obtained the orbital ground state, we are able to
determine the charge distribution at the Ti sites characterized
by the quadrupole moments. The tensor of the quadrupole moment per
one Ti position is given by
\begin{equation}
Q_{\alpha \beta }=\frac{2}{21}\left| e\right| \langle r^{2}
\rangle \langle 3l_{\alpha }l_{\beta }-6\delta _{\alpha \beta
}\rangle.
\end{equation}
Diagonalization of the tensors $Q_{\alpha \beta}/(|e|\langle
r^{2}\rangle)$ calculated for all four Ti positions yields the
same principal values equal to $Q_1=-0.520$, $Q_2=0.460$, and
$Q_3=0.060$, i. e. the charge distribution on the titanium ions is
the same in the local coordinate systems, which are just rotated
with respect to each other. The angles of rotations have been
calculated via the eigenvectors of the tensors $Q_{\alpha
\beta}$. The components of the unit vectors ($n_x$, $n_y$, $n_z$)
corresponding to the principal values -0.520 and 0.460 read
$\mathbf{n}_1=(0.815, 0.573, 0.086)$ and $\mathbf{n}_2=(-0.573,
0.746, 0.355)$ at the Ti1 site. For -0.520 (0.460) $n_y$ ($n_x$)
is reversed at the Ti3 and Ti4 sites, whereas $n_z$ is reversed at
the Ti2 (Ti2) and Ti4 (Ti3) sites.

It is interesting to know, how the spin is oriented with respect
to the quadrupole charge distribution. According to
neutron-scattering data \cite{Cwik2003,Reehuis2004} and
susceptibility measurements \cite{Meijer1999,Fritsch2002} the
effective magnetic moment per one Ti$^{3+}$ is about $\mu_{\rm
eff}\sim 0.6\mu_{B}$. The antiferromagnetically ordered moments
are aligned along the $c$--direction and weak ferromagnetism
shows up along the $b$--direction (in Pnma).\cite{Cwik2003} We
suggest that this can be explained as follows: Due to the
spin-orbit coupling the orientations of the titanium magnetic
moments are connected with the quadrupole ordering. If we assume
that the spin is aligned perpendicular to the $3d$--electron
charge-distribution plane, i. e. along $\mathbf{n}_{2}$, a
ferromagnetic alignement along the $b$--axis can result from the
$y$ component of $\mathbf{n}_2$, which is positive at all four Ti
places, and a $G$--type antiferromagnetic order along the
$c$--axis is favored as the sign of the $z$--component of
$\mathbf{n}_2$ changes between the Ti sites, correspondingly. As
neutron scattering detects the averaged magnetic moment of the
four inequivalent Ti places per unit cell with vice versa
twisting of the quadrupolar moments, the observed $\mu_{\rm
eff}\sim 0.6\mu_{B}$ is just the projection of the total magnetic
moments onto the $c$--direction.

\section{\label{chi}Magnetic susceptibility}

The LaTiO$_3$ single crystal, prepared by floating zone
melting,\cite{Lichtenberg1991} was essentially the same as used
previously for the thermal-expansion measurements described in
Ref.~\onlinecite{Hemberger2003}. The crystallographic axes were
determined from x-ray Laue pictures. Additional
neutron-diffraction experiments\cite{Reehuis2004} on the same
crystal revealed only a small twin-domain of about 5\% of the
crystal volume, hence the crystal can be regarded as practically
untwinned. The magnetization $M(T)$ was measured in a commercial
superconducting quantum interference device (SQUID) magnetometer
(MPMS5, {\it Quantum Design}), working in a temperature range
$1.8 \leq T \leq 400$~K and in magnetic fields up to $H=50$~kOe.

\begin{figure}
\includegraphics[width=7cm]{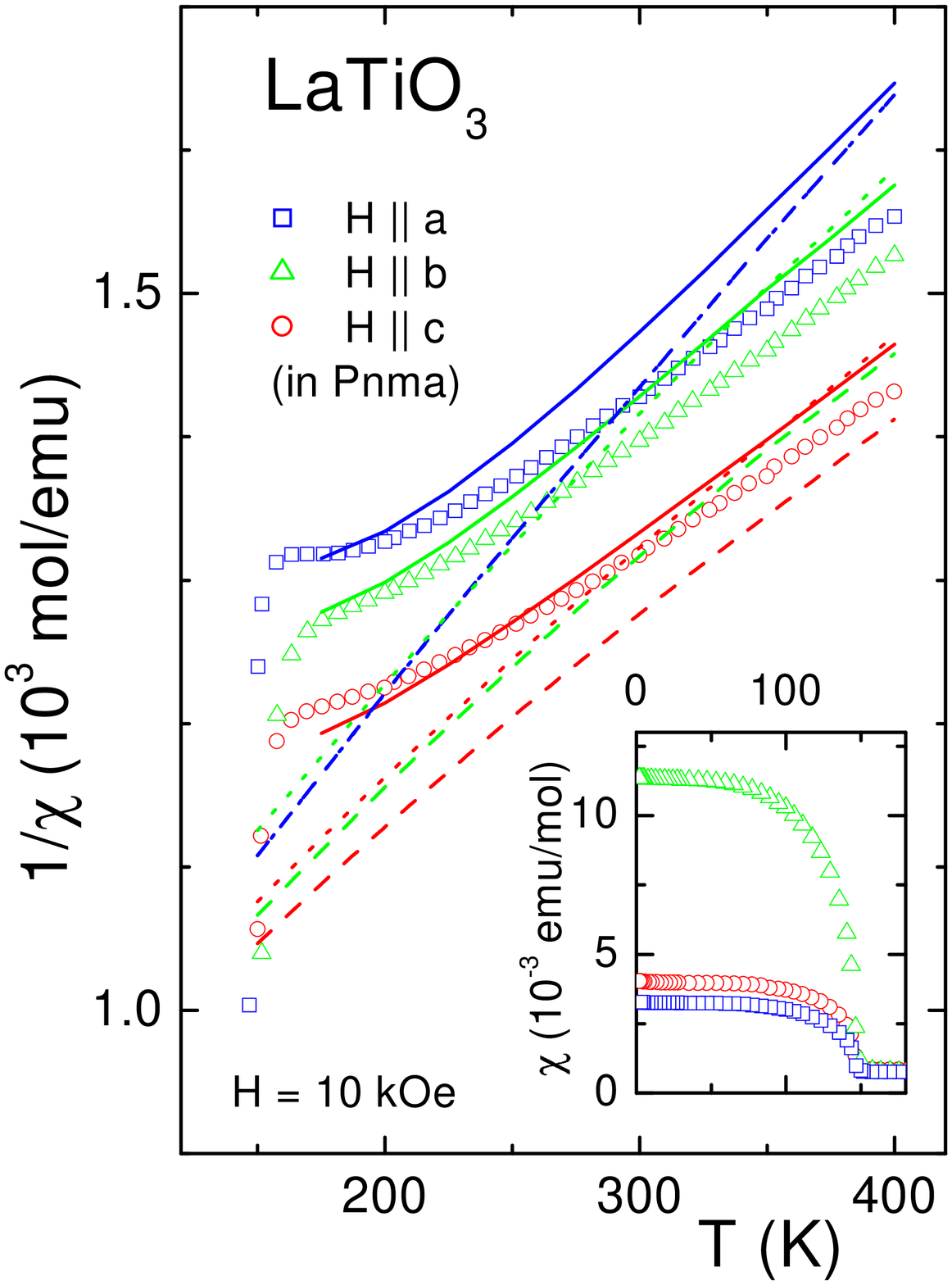}
\caption{Temperature dependence of the inverse susceptibility
$1/\chi(T)$ (Inset: $\chi(T)$ at low temperatures) in LaTiO$_3$
for an external field of $H=10$~kOe applied along the three
crystalographic axes $a$, $b$, and $c$ (Pnma). The fits indicated
by solid, dashed, and dotted lines are described in the text
(color online). \label{sus}}
\end{figure}

Figure~\ref{sus} shows the temperature dependence of the
susceptibility $\chi = M/H$ obtained from the LaTiO$_3$ single
crystal in an external field of $H=10$~kOe applied along the
three orthorhombic axes both below $T_{\rm N}$ (inset) and in
inverse representation in the paramagnetic regime (main frame).
The data have been corrected accounting for the diamagnetic
background of the sample holder, which was measured independently
for all three geometries. Below the N\'eel temperature $T_{\rm
N}=146$~K, one observes the evolution of a weak ferromagnetic
magnetization of about $0.02\mu_{\rm B}$ per formula unit with
its easy direction along the $c$--axis. The paramagnetic regime is
better visible in the inverse susceptibility with an
approximately linear increase above 200~K. Evaluation by a
Curie-Weiss behavior $N_{\rm A} \mu_{\rm eff}^2/3k_{\rm
B}(T+\Theta_{\rm CW})$, with $\mu_{\rm eff}^2= \mu_{\rm
B}^2g^2S(S+1)$ yields a Curie-Weiss temperature $\Theta_{\rm CW}
\approx 900$~K and an effective moment $\mu_{\rm eff} \approx
2.6\mu_{\rm B}$, which is strongly enhanced with respect to the
spin-only value of $1.73\mu_{\rm B}$. For an appropriate
evaluation we have to take into account the preceding
energy-level scheme derived from our CF analysis.

Including the external magnetic field, the perturbation
Hamiltonian is written as
\begin{equation}
V=-\mu_{\rm B} H_{\alpha }(k_{\alpha }l_{\alpha }+2s_{\alpha
}\mathbf{-}f_{\alpha }s_{\alpha })=-H_{\alpha}M_{\alpha}.
\end{equation}
Here the factors $f_{\alpha}$ take into account the molecular
field, which can be anisotropic for two reasons. The first one is
because of the anisotropic $g$--factors. The second one is due to
the anisotropy of the effective superexchange interaction between
the titanium spins, which we take in the form
$\underset{\alpha}\sum J_{ij}^{\alpha}s_{\alpha}^{i}s_{\alpha
}^{j}$. The parameters $J_{ij}^{\alpha}$ represent the effective
superexchange integrals, $\alpha =x,y,z$. In the crystal
structure around each Ti$^{3+}$ ion, there are two titanium ions
at a distance R$_{1}=3.958$~\AA, four titanium ions at
R$_{2}=3.971$~\AA, and 12 at a distance R$_{3}\approx 5.6$~\AA.
According to the neutron-scattering data \cite{Keimer2000}
$J_{1}^{\alpha }\approx J_{2}^{\alpha }\approx 180$~K for all
$\alpha$.

The molecular field approximation taking into account the six
nearest neighbors at distances $R_1$ and $R_2$ yields
\begin{equation}
f_{\alpha} = \frac{6J\langle s_{\alpha} \rangle \langle
k_{\alpha}l_{\alpha}+2s_{\alpha}\rangle}{k_{\rm B}T+6J\langle
s_{\alpha} \rangle^2} = \frac{C_{\alpha }}{T+\Theta_{\alpha}}.
\end{equation}
Note that in this approximation the ratios
$C_{\alpha}/\Theta_{\alpha}$ are independent on the exchange
coupling $J_{\alpha}$ and directly determined by the spin- and
orbital state as
\begin{equation}
\frac{C_{\alpha}}{\Theta_{\alpha}}=\frac{\langle\varepsilon_{1}\mid
k_{\alpha}l_{\alpha}+2s_{\alpha}\mid
\varepsilon_{1}\rangle}{\langle\varepsilon_{1}\mid s_{\alpha}\mid
\varepsilon_{1}\rangle}.
\end{equation}
The ratios $C_{x}/\Theta_{x}\approx 1.92$ and
$C_{y}/\Theta_{y}\approx 1.85$, and $C_{z}/\Theta_{z}\approx
1.80$, as calculated from the ground state assuming
$k_{\alpha}=1$, indicate again an only small contribution of the
orbital momentum $l_{\alpha}$ to the magnetic susceptibility. For
zero orbital momentum one would obtain
$C_{\alpha}/\Theta_{\alpha}=2$.

For $\alpha = z$ the paramagnetic part of the susceptibility can
be written as
\begin{equation}
\chi _{para}^{zz}=\frac{1}{Z}\underset{l}{\sum}
\langle\varepsilon_{l}\mid M_{z} \mid
\varepsilon_{l}\rangle^{2}\exp (-\varepsilon_{l}/k_{\rm B}T)
\end{equation}
where $Z=k_{\rm B}T \underset{l}{\sum} \exp
(-\varepsilon_{l}/k_{\rm B}T)$. The van Vleck like contribution
reads
\begin{eqnarray}\label{chiVV}
\chi _{\rm vv}^{zz}= 2\underset{l}{\sum}^{\prime
}\frac{\langle\varepsilon_{1}\mid M_{z}\mid
\varepsilon_{l}\rangle\langle\varepsilon_{l}\mid M_{z} \mid
\varepsilon_{1}\rangle}{\varepsilon_{l}-\varepsilon _{1}}.
\end{eqnarray}
The cases $\alpha =x,y$ can be written analogously. In addition,
one has to take into account the diamagnetic susceptibility. It
can be estimated from the ionic susceptibilities\cite{Landolt1976}
(given in $10^{-6}$~emu/mol) of Sr$^{2+}$ (-15), which is
isoelectronic to La$^{3+}$, Ti$^{3+}$ (-9), and O$^{2-}$ (-12) as
$\chi_{\rm dia}=-6 \cdot 10^{-5}$~emu/mol.

In Fig.~\ref{sus} the theoretical description of the data has
been performed in three steps, as illustrated by the three groups
of dashed, dotted, and solid lines, respectively. In the first
step (dashed lines) the exchange coupling is assumed to be
isotropic and used as the only fit parameter $J_{\alpha}=J$. The
reduction factors have been kept fixed at $k_{\alpha}=1$. With
$J=200$~K in good agreement with the results of neutron
scattering, one achieves already a reasonable description of the
susceptibility. It is remarkable that absolute value and
anisotropy are very well reproduced by this straight-forward
calculation.

In the second step, we allowed a variation of the covalency
parameters $k_{\alpha}$. With the same exchange constant of 200~K
and $k_x=1$, $k_y=0.88$, and $k_z=0.95$ (dotted lines) the
description of the relative splitting of the susceptibilities
between the different axes is improved, but the curvature is
still not reproduced. Nevertheless, the obtained covalency
parameters match the values typically observed for Ti$^{3+}$
ions.\cite{Abragam1970} The resulting ratios $C_{\alpha
}/\Theta_{\alpha }$ change only slightly $C_{x}/\Theta_{x}\approx
1.92$ and $C_{y}/\Theta_{y}\approx 1.87$, and
$C_{z}/\Theta_{z}\approx 1.81$ with respect to $k_{\alpha}=1$.

Finally in the third step, the solid lines show the fit of the
experimental data using in addition the values $C_{\alpha}$ and
$\Theta_{\alpha}$ as adjustable parameters. From fitting we have
got $C_{x}/\Theta_{x}=2.45$, $C_{y}/\Theta_{y}=2.29$, and
$C_{z}/\Theta_{z}=2.14$. These deviations from the
nearest-neighbor isotropic molecular-field results can be
considered as a hint for an spin-orbit dependent exchange like
$J(s_{i}s_{j}) l_{\alpha}^{i} l_{\beta}^{j}$ between the titanium
ions. In principle such kind of operators are known and have been
discussed in a number of papers
\cite{Drillon1982,Leuenberger1984,Ratikin1995,Borras2001} in
application to the susceptibility of the dimer
[Ti$_{2}$Cl$_{9}$]$^{-3}$ and \textit{a-priory} cannot be
discarded for LaTiO$_{3}$. Another influence, which to our opinion
cannot be excluded as well, is the next nearest neighbor
interaction between the titanium ions. Obviously this question
should be addressed to further analysis, when more experimental
information will be obtained. However, both types of mentioned
interactions can produce the corrections of a few percent, but we
believe that the essential physics of the temperature dependence
of magnetic susceptibility and orbital ordering will be the same
as described above.

Note that the factor $f_{\alpha}$ is quite large and, therefore,
according to Eq.~\ref{chiVV} $\chi_{\rm vv}^{\alpha \alpha }$ is
dependent on temperature. This fact to our knowledge has not been
pointed out in literature. We think that this situation should be
quite general for other titanium compounds as well as for
vanadium oxides.

\section{Conclusions}

In summary the energy splitting and wave functions of the
Ti$^{3+}$ $3d^1$--electron state have been calculated for
LaTiO$_{3}$ due to the crystal field including spin-orbit
coupling and Ti--O exchange. From the derived orbital ground state
we have estimated the quadrupole moments at the Ti sites and have
deduced the charge-distribution picture for the $3d$--electrons in
the crystallographic coordinate system. Based on the orientation
of the quadrupolar tensor, it is possible to suggest an
explanation for the low value of the ordered moment, observed in
the antiferromagnetic state. The straight-forward calculation of
the paramagnetic susceptibility yields the correct anisotropy,
which we measured in an untwinned LaTiO$_3$ single crystal.

\section{Acknowledgments}

We thank Dana Vieweg for performing the SQUID measurements. This
work is supported by the German Bundesministerium f\"ur Bildung
und Forschung (BMBF) under the contract No. VDI/EKM 13N6917, by
the Deutsche Forschungsgemeinschaft (DFG) via
Sonderforschungsbereich SFB 484 (Augsburg), by the Russian RFFI
(Grant 03 02 17430) and partially by CRDF (BRHE REC007).

\end{document}